\documentclass[10pt,conference]{IEEEtran}

\usepackage[noadjust]{cite}
\usepackage{graphicx,color}
\usepackage[dvipsnames]{xcolor}
\usepackage{amsmath,amsbsy,amsfonts,amssymb}
\usepackage{mathrsfs}
\usepackage{mathtools}
\usepackage{flushend}
\usepackage{amsthm}  % for \newtheorem
\usepackage{amssymb}
\usepackage[mathscr]{eucal}
\usepackage{bm}
\usepackage{algorithm}
\usepackage{algpseudocode}
\usepackage[percent]{overpic}
\usepackage{pgfplots}

\usepackage{graphicx}
\usepackage{tikz}
\usetikzlibrary{arrows.meta} % punte moderne
\definecolor{quadTL}{HTML}{9A82E0} % top-left (darker lavender)
\definecolor{quadTR}{HTML}{8A949E} % top-right (darker blue-grey)
\definecolor{quadBL}{HTML}{B58383} % bottom-left (darker pink)
\definecolor{quadBR}{HTML}{9E9157} % bottom-right (darker beige)
\usepackage{scalerel}
\usepackage{adjustbox}
\usepackage{float}
\usepackage{comment}

\usepackage[normalem]{ulem}
\usepackage{braket}
\usepackage[font=footnotesize]{caption}   
\usepackage{placeins}

\usepackage{xcolor} % Necessario per i colori

\usepackage{subcaption}
\usepackage[hidelinks]{hyperref}

\usepackage{orcidlink}

\usepackage{amssymb}

\algnewcommand\Input{\item[\textbf{Input:}]}
\algnewcommand\Output{\item[\textbf{Output:}]}

\usepackage[acronym]{glossaries}

\IEEEoverridecommandlockouts % To allow \thanks{} command to work

\begin{document}

\newcommand{\weight}{w}
\newcommand{\indexchecknode}{j}
\newcommand{\indexvariablenode}{i}
\newcommand{\variablenode}[1]{\ensuremath{v_{#1}}}
\newcommand{\checknode}[1][]{\ensuremath{c_{#1}}}
\newcommand{\arrowtoright}{i \rightarrow j}
\newcommand{\LLRsvector}[2]{\ensuremath{\mathbf{\Gamma}_{{{#1}\to {#2}}}}}
\newcommand{\LLR}{\Gamma}
\newcommand{\LLRsvectorcomponent}[2]{\ensuremath{\Gamma_{#1\to #2}}}
\newcommand{\LLrcomponentbold}[2]{\ensuremath{\Lambda_{#1\to #2}}}
\newcommand{\LLRcomponent}[1]{\ensuremath{\Lambda_}{#1}}
\newcommand{\error}[1][]{\ensuremath{e_{#1}}}
\newcommand{\beliefoperator}[1][]{\ensuremath{\phi_{#1}}}
\newcommand{\stabilizermatrix}[1]{G_{#1}}
\newcommand{\stabilizer}{G}
\newcommand{\numberstabilizer}{m}
\newcommand{\numberqubits}{n}
\newcommand{\messagechecknodes}{\Delta}
\newcommand{\arrowtoleft}{i \leftarrow j}
\newcommand{\syndromecomponent}{s}
\newcommand{\neighboringvn}{\mathcal{N}}
\newcommand{\neighboringcn}{\mathcal{M}}
\newcommand{\weightvn}[1][]{\ensuremath{ w_{v, #1}}}
\newcommand{\indexiterations}{l}
\newcommand{\weightcn}[1][]{\ensuremath{ w_{c, #1}}}
\newcommand{\logqubits}{k}
\newcommand{\physicalqubit}{n}
\newcommand{\paulioperators}{P}
\newcommand{\codedistance}{d}
\newcommand{\prob}{p}
\newcommand{\stabilizergroup}{G}
\newcommand{\xbit}{x}
\newcommand{\zbit}{z}
\newcommand{\checkmatrix}{H}

\newcommand{\FL}{\textcolor{blue}}

\title{\huge Neural Belief‑Matching Decoding for Topological Quantum Error Correction Codes %with Convolutional Weight‑Sharing %Neural Belief-Matching with Convolutional Architectures for Quantum Error Correction
}

\author{
  \IEEEauthorblockN{Luca Menti\orcidlink{0009-0001-2222-9223}, 
                    Francisco L\'azaro\orcidlink{0000-0003-0761-7700}}
  \IEEEauthorblockA{Institute of Communications and Navigation\\
  German Aerospace Center (DLR) \\
  email: \{Luca.Menti, Francisco.LazaroBlasco\}@dlr.de}

\vspace{-0.5cm}

\thanks{This work is part of the HESC lighthouse project within Munich Quantum Valley initiative and is supported by the Bavarian state government with funds from the Hightech Agenda Bavaria.} 
\thanks{Luca Menti's research was funded by a scholarship awarded by the German Academic Exchange Service (DAAD).}
\thanks{© 2026 IEEE. Personal use of this material is permitted. Permission from IEEE must be obtained for all other uses, including reprinting/republishing this material for advertising or promotional purposes, collecting new collected works for resale or redistribution to servers or lists, or reuse of any copyrighted component of this work in other works.}
%}
}

%\vspace{-15mm}
\date{}
\maketitle

\thispagestyle{empty}
\pagestyle{empty}
\newacronym{qec}{QEC}{quantum error correction}
\newacronym{bp4}{BP4}{quaternary BP}
\newacronym{qecc}{QECC}{quantum error correction codes }
\newacronym{qldpc}{QLDPC}{Quantum Low-Density Parity Check }
\newacronym{qtc}{QTC}{Quantum Turbo Codes }
\newacronym{mwpm}{MWPM}{minimum weight perfect matching }
\newacronym{uf}{UF}{union find }
\newacronym{bp}{BP}{belief-propagation}
\newacronym{nn}{NN}{neural networks}
\newacronym{ml}{ML}{machine-learning}
%\newacronym{nbp}{NBP}{Neural Belief Propagation}
\newacronym{nbp4}{NBP}{quaternary neural belief-propagation}
\newacronym{cnn}{CNN}{convolutional neural networks}
\newacronym{cnbp}{convolutional NBP}{quaternary convolutional neural belief-propagation}
\newacronym{pcm}{PCM}{parity-check matrix}
\newacronym{ocm}{OCM}{overcomplete check matrix}
\newacronym{css}{CSS}{Calderbank–Shor–Steane}
\newacronym{cn}{CN}{check node}
\newacronym{vn}{VN}{variable node}
\newacronym{sgd}{SGD}{stochastic gradient descent}
\newacronym{onbp}{$\text{RNBP}$}{NBP with overcomplete check matrix}
\newacronym{ocnbp}{convolutional RNBP}{convolutional NBP with overcomplete check matrix}
\newacronym{nbm4}{NBP-matching}{quaternary neural belief-matching}
\newacronym{nbuf4}{NBP-Find}{quaternary Neural Belief-Find}
\newacronym{cbm4}{Convolutional NBP-matching}{quaternary convolutional  belief-matching}
\newacronym{cbuf4}{Convolutional NBP-Find}{Quaternary Convolutional  Belief-Find}
\newacronym{ler}{LER}{logical error rate}

%%% TO DISCUSS

%- authors for suppressing quantum error

%- index terms, funding acknoledmenr

%-

\begin{abstract}
Quantum error correction (QEC) is critical for scalable fault-tolerant quantum computing. 
Topological codes, such as the toric code, offer hardware-efficient architectures but their Tanner graphs contain many girth-4 cycles that degrade the performance of \gls{bp} decoding. 
For this reason, BP decoding is typically followed by a more complex second stage decoder such as minimum-weight perfect matching. 
These combined decoders achieve a remarkable performance, albeit at the cost of increased complexity.
In this paper we propose two key improvements for the decoding of toric code. %topological codes. CHECK With Fran
The first one is replacing the BP decoder by a neural BP decoder, giving rise to the neural belief-matching  decoder which substantially decreases the average decoding complexity. 
The main drawback of this approach is the high cost associated with the training of the neural BP decoder. 
To address  this issue, we impose a convolutional architecture on the neural BP decoder, enabling weight sharing across the spatially homogeneous structure of the code’s factor graph. This design allows a model trained on a modest‑size topological code to be directly transferred to much larger instances, preserving decoding quality while dramatically lowering the training burden.
Our numerical experiments on toric‑code lattices of various sizes demonstrate that this technique does not result in a noticeable loss in performance.
\end{abstract}


% here add latex files containing your sections

\section{Introduction}
\Gls{qec} schemes are essential to realizing fault-tolerant quantum computation. Since the early breakthroughs, the field has grown dramatically, and today we have multiple families of \gls{qecc}, including topological codes \cite{kitaev2003fault}, \gls{qldpc} codes \cite{mackay2004sparse} and \gls{qtc} \cite{poulin2009quantum}. 
Topological codes offer several key advantages. They operate using only local, nearest‐neighbor interactions and every stabilizer measurement involves only
a few adjacent qubits. For these reasons, they are regarded as the preferred \gls{qec} solution for superconducting-qubit platforms \cite{fowler2012towards, terhal2015quantum}.
%
%Moreover, by implementing the code on an open, planar sheet, logical qubits are embedded in a flat, rectangular array where only the boundary conditions differ. This planar configuration meshes naturally with chip-based qubit platforms and supports tiled extensions to achieve arbitrarily large logical distances \cite{fowler2012surface}.
%\\

Among the many different decoding algorithms that have been proposed for topological codes, the most widely used is the \gls{mwpm} decoder  \cite{dennis2002topological, edmonds1965paths} which provides an excellent error-correction capability, albeit at the cost of a large decoding complexity. In particular its worst-case runtime is $O(n^3\text{log}(n))$, with  $n$ being the number of (physical) qubits \cite{dennis2002topological,higgott2022pymatching,edmonds1965paths}.
%
%Recently, the \gls{uf} decoding algorithm \cite{delfosse2021almost,huang2020fault} was proposed, which provides a slightly lower error-correction performance while exhibiting lower decoding complexity.

\gls{mwpm} is optimal\footnote{it returns the most likely error} under independent bit- and phase-flip errors. However, it is suboptimal under the depolarizing channel due to the presence of correlated $\boldsymbol{X}\!-\!\boldsymbol{Z}$ (i.e., $\boldsymbol{Y}$) errors.

A very appealing decoding algorithm is \gls{bp}, which exhibits a low decoding complexity, is easily parallelized in hardware implementations and can  inherently deal with $\boldsymbol{X}\!-\!\boldsymbol{Z}$ (i.e., $\boldsymbol{Y}$) errors. However, the performance of topological codes under \gls{bp} decoding is very poor due to the large number of girth-4 cycles in their Tanner graph.
%Decoders for these codes belong mainly to two classes: graph-theory-based decoders such as the \gls{mwpm} decoder  \cite{dennis2002topological, edmonds1965paths}, and message-passing decoders, notably \gls{bp}. \gls{mwpm} is widely used for topological codes but its main drawback is high decoding latency. By contrast, the latency of \gls{bp} scales linearly with the block length and the number of iterations; however, topological codes unavoidably introduce girth-4 cycles in their Tanner graphs by construction, which degrade \gls{bp} performance.\\
%While \gls{bp} decoders can perform very well on Tanner graphs whose girth is at least 6, topological codes unavoidably introduce girth-4 cycles in the Tanner graph by construction, which deteriorate \gls{bp} performances.\\
%

%% should we remove uf since we are considering just matching?

An effective decoding approach is using \gls{bp} as a first stage decoding and feeding its output into a second stage \gls{mwpm} decoder, giving rise to the belief-matching decoder \cite{higgott2023improved}. 
This technique not only improves the error-correcting performance but  also yields a lower average decoding complexity, since the second stage decoder only needs to be used when \gls{bp} fails to converge.

%Notable approaches to address this challenge include the Belief-Matching and Belief-Find decoders \cite{higgott2023improved}. In this framework, \gls{bp} initially refines per-qubit error probabilities during preprocessing. If \gls{bp} fails to converge, its belief estimates are used to reweight the edges of the matching graph, after which \gls{mwpm} or weighted \gls{uf} \cite{delfosse2021almost,huang2020fault} performs the final decoding. By integrating \gls{bp}'s adaptive probability updates with \gls{mwpm}/\gls{uf}'s optimization capabilities, these decoders achieve significantly higher accuracy than either method alone.\\ 

Another interesting approach to improve the performance of \gls{bp} decoding is neural \gls{bp} \cite{nachmani2016learning}, which reinterprets the decoder as a neural network and adds trainable multiplicative weights to the exchanged messages. This technique was originally introduced for classical error correcting codes but has also been proposed for \gls{qec} codes \cite{liu2019, miao2025quaternary}.
The main drawback of this technique is that training becomes increasingly challenging (in terms of computational and memory resources) for larger code distances. This problem becomes even more acute when considering circuit-level noise. Moreover, its overall performance remains inferior to that of the belief-matching
%and Belief-Find 
under the phenomenological depolarizing noise model, as it is shown in this paper.

The contribution of this paper is twofold. First, we propose replacing the initial \gls{bp} stage in belief-matching by neural \gls{bp}. As we show in our numerical results, this 
%does improve the error-correcting performance of belief-matching or belief-union, but it 
drastically reduces the average decoding complexity. In particular, the number of times that the \gls{mwpm} decoder is called is reduced by up to 4 orders of magnitude.
The second contribution is an approach to dramatically reduce the training cost of neural \gls{bp} for topological codes. In particular, we exploit the translational symmetry of the two dimensional lattice and introduce a convolutional structure in the neural \gls{bp} decoder. This approach enables to train our decoder on a small topological code, and reuse the obtained weights for larger codes. Our results indicate that this approach is effective and does not incur any noticeable loss in performance.

\section{Background}
\label{sec:Background}

\subsection{Preliminaries}
%Quantum error correction encodes $\logqubits$ logical qubits into $\numberqubits$ physical qubits so that, despite noise, the logical qubits behave as if they were error‐free.

%\FL{shorter, look at Valentini}

The $\numberqubits$‑qubit Pauli group (phases omitted) is $
\boldsymbol{\mathcal{P}}
=\bigl\{\boldsymbol{P}_1\otimes\cdots\otimes \boldsymbol{P}_{\numberqubits}\,\bigm|\,\boldsymbol{P}_i\in\{\boldsymbol{I},\boldsymbol{X},\boldsymbol{Y},\boldsymbol{Z}\}\bigr\}
$. The weight $w $ of a Pauli operator $\boldsymbol{P}=\boldsymbol{P}_1\otimes\cdots\otimes \boldsymbol{P}_{\numberqubits}$ is the number of non‑identity
tensor factors. A $[[\numberqubits,\logqubits,\codedistance]]$ stabilizer code is defined by an abelian subgroup $\boldsymbol{\mathcal{S}}\subset\boldsymbol{\mathcal{P}}$ generated by $\numberstabilizer=\numberqubits-\logqubits$  independent, commuting Pauli checks; 
%The stabilizer generators are realized as projective measurements, implemented via ancillary qubits, that leave the encoded state invariant.
it encodes $\logqubits$
logical qubits into $\numberqubits$ physical qubits, has  minimum distance $\codedistance$ and  can correct any error acting up to  $\lfloor (\codedistance-1)/{2} \rfloor$ qubits.  The code space $\mathcal{C}$ is defined as the set of states $\ket{\psi}$ for which $\boldsymbol{S}\ket{\psi} = \ket{\psi}$ holds for all $\boldsymbol{S} \in \boldsymbol{\mathcal{S}}$ \cite{gottesman1997stabilizer,
fuentes2021degeneracy,brun2006correcting}.  \\
Writing a Pauli as $\boldsymbol{\mathcal{P}} = \bigotimes_{i=1}^{\numberqubits}\boldsymbol{X}^{x_i}\boldsymbol{Z}^{z_i}$ maps it to the binary vector $(\mathbf{x}\mid\mathbf{z})\in\mathbb{F}_2^{2\numberqubits}$. Two Paulis commute iff their symplectic product vanishes, $\sum_{i=1}^{\numberqubits}(x_i b_{Z,i} + z_i b_{X,i}) \equiv 0 \pmod 2.$ Hence, a stabilizer code is described by a binary parity‑check matrix $ \boldsymbol{\checkmatrix} 
%= \bigl(\boldsymbol{\checkmatrix} _X \mid \boldsymbol{\checkmatrix} _Z\bigr)
\in\mathbb{F}_2^{(2\numberqubits-\logqubits)\times2\numberqubits}$.\\
For \gls{css} codes $\boldsymbol{\checkmatrix}$ block-diagonalizes as  
$
\boldsymbol{\checkmatrix}  = \begin{pmatrix}
\boldsymbol{\checkmatrix}_X & 0\\
0    & \boldsymbol{\checkmatrix}_Z
\end{pmatrix}
$
\cite{gottesman1997stabilizer}.
%The toric code is a prominent example.\\

The Pauli operators can also be represented as elements of the finite field $\mathbb{F}_4 = \{0, 1, \omega, \bar{\omega}\}$.
%, where $\omega$ is a primitive element satisfying $\omega^2 = \omega + 1$ and $\bar{\omega} = \omega^2$. 
Each Pauli string is encoded via the mapping $(x_i, z_i) \mapsto \alpha_i = x_i \omega + z_i \bar{\omega}$, where $(x_i, z_i) \in \{0, 1\}^2$ denotes the binary \(\boldsymbol{X}\)- and \(\boldsymbol{Z}\)-type coordinates. This transformation converts the binary check matrix $\boldsymbol{H}$ into a quaternary stabilizer matrix $\boldsymbol{G} \in \mathbb{F}_4^{(2n - \log_2 n) \times n}$.%, which simultaneously encodes all Pauli-type constraints for $n$ qubits.

%
%To handle all Pauli types simultaneously, one often works over $\mathbb{F}_{4}=\{0,1,\omega,\bar\omega\}$.  By mapping each pair $(x_i,z_i)$ to the field element $ \alpha_i = x_i\,\omega + z_i\,\bar\omega,$ the binary check matrix $\boldsymbol{\checkmatrix} $ becomes a quaternary matrix $\boldsymbol{\stabilizergroup}\in\mathbb{F}_{4}^{(2\numberqubits-\logqubits)\times\numberqubits}.$
%The symplectic condition then translates into the trace-Hermitian inner-product constraint $\langle a,b\rangle = \sum_{i=1}^{\numberqubits}\mathrm{tr}\bigl(a_i\,\overline{b_i}\bigr)=0,$ where $\mathrm{tr}(\omega)=\mathrm{tr}(\bar\omega)=1$ and $\mathrm{tr}(0)=\mathrm{tr}(1)=0$. 

The normalizer of the stabilizer group $\mathcal{S}$, denoted $\mathcal{N}(\mathcal{S})$, consists of all Pauli operators commuting with every element of $\mathcal{S}$.  $\mathcal{N}(\mathcal{S})$ is generated by $2n - m$ independent operators, where $m$ is the number of stabilizer generators.  The matrix $\boldsymbol{\stabilizergroup}^\perp \in \mathbb{F}_4^{(2n - m) \times n}$ contains the generators of $\mathcal{N}(\mathcal{S}) \setminus \mathcal{S}$ as rows, representing the logical operators. The code's minimum distance $d$ equals the smallest weight of any such logical operator.

For an error $\boldsymbol{E}\in\boldsymbol{\mathcal P}$, measuring all
stabilizer generators yields the syndrome
$\boldsymbol s\in\{\pm1\}^{m}$.  
Let $\boldsymbol{\hat{E}}$ be the error estimated provided by the decoder. Error correction is successful iff $\boldsymbol{E}\boldsymbol{\hat{E}} \in \boldsymbol{\mathcal{S}}$.
%The decoder proposes a correction
%$\boldsymbol{\hat{E}}$; the error is corrected iff
%$\boldsymbol{E}\boldsymbol{\hat{E}} \in \boldsymbol{\mathcal{S}}$.\\
%In this work we consider the phenomenological depolarizing model: each qubit suffers an $\boldsymbol{X}$, $\boldsymbol{Y}$, or $\boldsymbol{Z}$ error independently with probability $\epsilon/3$, so the physical error rate is $\epsilon$.
 In this work we adopt the phenomenological depolarizing noise model, in which each qubit independently undergoes a Pauli $\boldsymbol{X}$, $\boldsymbol{Y}$, or $\boldsymbol{Z}$ error with probability $\epsilon/3$, so that the overall physical error rate is $\epsilon$.

 \subsection{Toric code}
\begin{figure}[H]
    \centering
    % riduco larghezza al 60% della colonna
    \includegraphics[width=0.6\columnwidth]{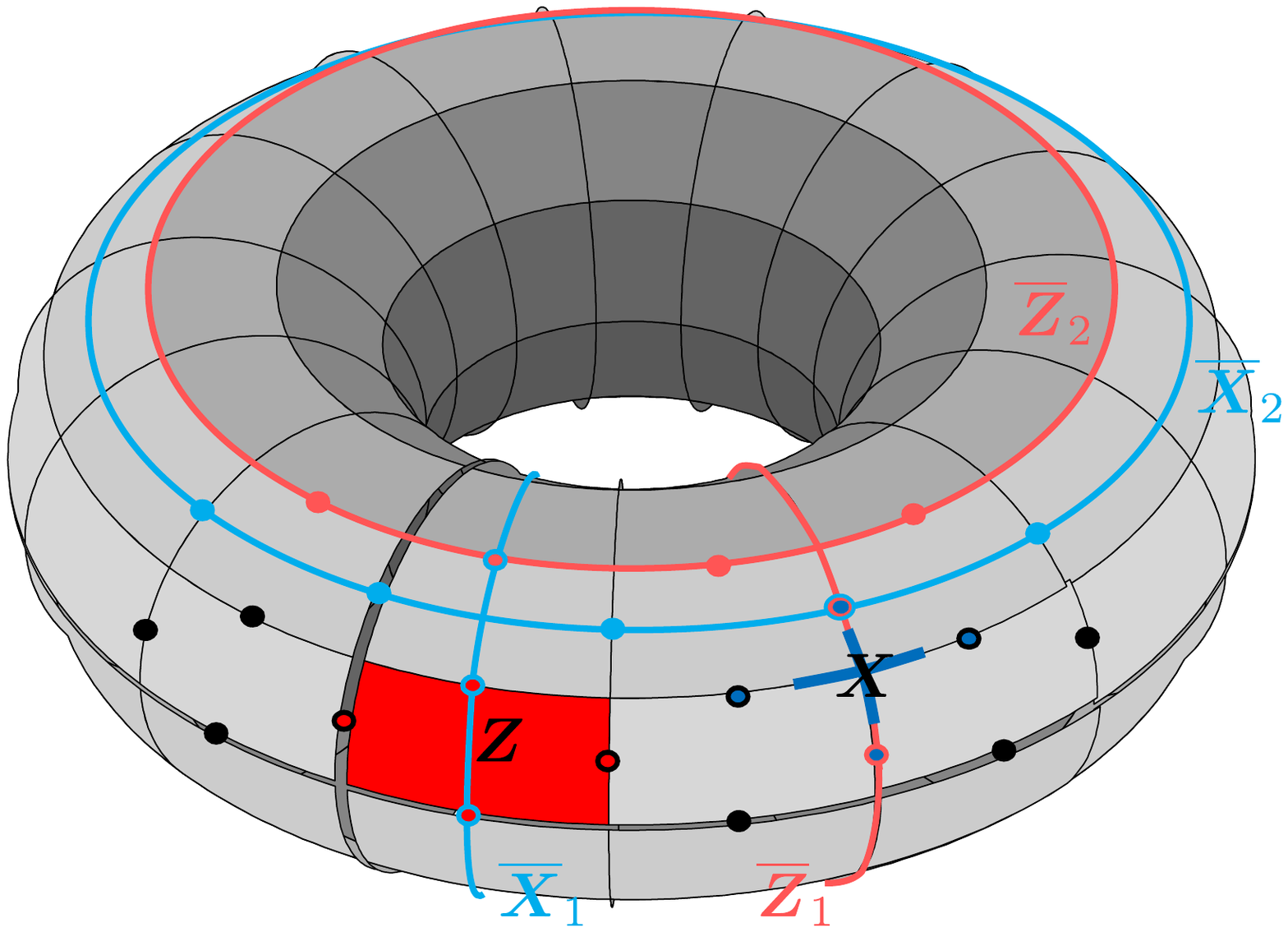}
        % coordinate (x,y) in percento rispetto all'immagine

\caption{Illustration of the toric code. Edge qubits (dots) are measured by weight‑4 stabilizers \(\boldsymbol{X}\)-type vertex checks (blue) and \(\boldsymbol{Z}\)‑type plaquette checks  (red). Light‑blue and light‑red non‑contractible strings denote the logical operators \(\overline{\boldsymbol{X}}_{1,2}\)  and \(\overline{\boldsymbol{Z}}_{1,2}\)  that wrap the two cycles of the torus.}

    \label{fig:im1}
\end{figure}

 The toric code is a topological stabilizer code on a torus encoding two logical qubits \cite{kitaev2003fault}. A \(\bigl[ \bigl[\numberqubits=2\codedistance^2,\,\logqubits=2,\,\codedistance\bigr] \bigr]\) toric code is obtained by placing qubits on the edges of a \(\codedistance \times \codedistance\) square lattice with periodic boundary conditions. The stabilizer group is generated by weight-4 vertex and plaquette checks: vertex operators act as Pauli-\(\boldsymbol{X}\) on the four edges incident to each vertex, and plaquette operators act as Pauli-\(\boldsymbol{Z}\) on the four edges surrounding each face. A \(\codedistance \times \codedistance\) toric code therefore has \(\codedistance^{2}\) \(\boldsymbol{X}\)–type and \(\codedistance^{2}\) \(\boldsymbol{Z}\)–type stabilizers, with one of each type redundant due to global constraints on the torus. A schematic of the toric code is shown in Fig. \ref{fig:im1}.

By multiplying adjacent vertex/plaquette operators, one obtains two weight-6 stabilizers for each vertex/plaquette operator.
Together with the \(\numberqubits\) independent weight-4 checks, these define the overcomplete \(3\numberqubits \times \numberqubits\) check matrix used in this work\footnote{
As shown in \cite{miao2025quaternary}, augmenting the original full-rank parity-check matrix with such redundant rows significantly improves \gls{bp} and quaternary neural belief-propagation decoding performance on toric codes.}.

\captionsetup{font=footnotesize}

%\vspace{-0.5cm}

%A toric code has $\codedistance^{2}$ $\boldsymbol{X}$ stabilizers of weight 4 from the vertex operators and $\codedistance^{2}$ $\boldsymbol{Z}$ stabilizers of weight 4 from the plaquette operators. Among them, one $\boldsymbol{X}$ stabilizer and one $\boldsymbol{Z}$ stabilizer are redundant. Moreover, the topological structure of the toric code allows to construct 2$\numberqubits$ redundant weight-6 $\boldsymbol{X}$ stabilizer and 2$\numberqubits$ redundant weight-6 $\boldsymbol{Z}$ stabilizer for every toric code. Combined with the $\numberqubits$ weight-4 stabilizers, we obtain an \gls{ocm} of size $3\numberqubits \times \numberqubits$ for toric codes \cite{miao2025quaternary}, which we use in this work. The \gls{ocm} are  check matrices with redundant rows in addition to the original full-rank check matrix,  and it has been shown that their application leads to a significant improvement in decoding performance of toric codes \cite{miao2025quaternary}.

%\section{Decoders}

\subsection{Quaternary belief-propagation}
We use the log-domain quaternary \gls{bp} decoder of \cite{ostrev2024classical} on the Tanner graph defined by $\boldsymbol{\stabilizergroup}$. The graph has variable nodes (VNs) $\{\variablenode{\indexvariablenode} \mid \indexvariablenode=1,\dots,n\}$ (error symbols) and check nodes (CNs) $\{\checknode[\indexchecknode] \mid \indexchecknode=1,\dots,m\}$ (syndrome bits). An edge connects $\variablenode{\indexvariablenode}$ and $\checknode[\indexchecknode]$ iff $\boldsymbol{\stabilizergroup}_{\indexvariablenode\indexchecknode}\neq 0$, and the edge is labeled by $\boldsymbol{\stabilizergroup}_{\indexvariablenode\indexchecknode} \in \mathbb{F}_{4}$. Given a measured syndrome $\boldsymbol{s}\in\{0,1\}^m$, \gls{bp} iteratively exchanges log-messages along the edges to approximate the posterior marginal distribution of the error symbols (variable nodes). Compared to using two separate binary \gls{bp} decoders for $\boldsymbol{X}$ and $\boldsymbol{Z}$ checks, this quaternary decoder explicitly accounts for the correlation between $\boldsymbol{X}$ and $\boldsymbol{Z}$ errors induced by the depolarizing channel.\\
For each variable node $\variablenode{\indexvariablenode}$ we initialize the log-likelihood ratio (LLR) vector $\LLRsvector{i}{j}$ as
$
\mathbf{\LLRcomponent}_{\indexvariablenode}
=
\bigl(
\LLRcomponent{\indexvariablenode}^{(1)} \;
\LLRcomponent{\indexvariablenode}^{(\omega)} \;
\LLRcomponent{\indexvariablenode}^{(\bar\omega)}
\bigr) \in \mathbb{R}^3,
$
with
$
\LLRcomponent{\indexvariablenode}^{(\zeta)}
=
\ln \left(
\frac{P(\error[\indexvariablenode]=0)}{P(\error[\indexvariablenode]=\zeta)}
\right)
=
\ln \left(
\frac{1-\epsilon}{\epsilon/3}
\right)
$.
%where $\zeta \in \mathbb{F}_{4} \setminus \{0\}$ and $\epsilon$ is the physical error probability.\\
For each nonzero $\eta \in \boldsymbol{\stabilizergroup}$, define
$\beliefoperator[\eta] : \mathbb{R}^3 \to \mathbb{R}$, mapping an LLR vector to the scalar LLR of the binary variable \(\langle \error[\indexvariablenode], \eta \rangle\) (inner product over $\mathbb{F}_2$):\\
$$
\beliefoperator[\eta](\mathbf{\LLRcomponent}_{\indexvariablenode})
=
\ln\left(
\frac{P (\langle \error[\indexvariablenode], \eta \rangle=0)}
     {P (\langle \error[\indexvariablenode], \eta \rangle=1)}
\right)
=
\ln \left(
\frac{1+ e^{-\LLRcomponent{\indexvariablenode}^{(\eta)}}}
     {\sum_{\zeta \neq 0 , \zeta \neq \eta} e^{-\LLRcomponent{\indexvariablenode}^{(\zeta)}}}
\right).
$$\\
The initial scalar VN-to-CN messages  are computed as
$\beliefoperator[\arrowtoright] := \beliefoperator[{\stabilizermatrix{\indexchecknode, \indexvariablenode}}](\LLRsvector{\indexvariablenode}{\indexchecknode})$
where $\arrowtoright$ denotes the message from VN $\variablenode{\indexvariablenode}$ to CN $\checknode[\indexchecknode]$. The outgoing messages from CN $\checknode[\indexchecknode]$ are
\begin{equation}
\messagechecknodes_{\arrowtoleft}
=
(-1)^{\syndromecomponent_\indexchecknode}
\cdot
2\tanh^{-1}
\left(
\prod_{\indexvariablenode' \in \neighboringvn(\indexchecknode) \setminus \{\indexvariablenode\}}
\tanh \frac{\beliefoperator[\indexvariablenode' \rightarrow \indexchecknode]}{2}
\right),
\label{eq: message cn}
\end{equation}
%\vspace{-2mm}
where $\neighboringvn(\indexchecknode)$ denotes the neighboring VNs of $\checknode[\indexchecknode]$.\\
In the VN update, we compute
$
\LLRsvector{\indexvariablenode}{\indexchecknode}
=
\bigl(
\LLRsvectorcomponent{\indexvariablenode}{\indexchecknode}^{(1)} \;
\LLRsvectorcomponent{\indexvariablenode}{\indexchecknode}^{(\omega)} \;
\LLRsvectorcomponent{\indexvariablenode}{\indexchecknode}^{(\bar\omega)}
\bigr)
$
with
\begin{equation}
  \LLRsvectorcomponent{\indexvariablenode}{\indexchecknode}^{(\zeta)}
  =
  \LLRcomponent{\indexvariablenode}^{(\zeta)}
  +
  \sum_{\substack{
      \indexchecknode' \in \neighboringcn(\indexvariablenode)\setminus\{\indexchecknode\} \\
      \langle \zeta,\;\stabilizermatrix{\indexchecknode',\indexvariablenode}\rangle = 1
  }}
  \messagechecknodes_{\arrowtoleft'},
  \label{eq:message1}
\end{equation}
for all \(\zeta \in \mathbb{F}_{4} \setminus \{0\}\), where \(\neighboringcn(\indexvariablenode)\) denotes the neighboring CNs of \(\variablenode{\indexvariablenode}\). The outgoing VN-to-CN messages are then \\
$
\beliefoperator[\arrowtoright]
=
\beliefoperator[\stabilizermatrix{\indexchecknode, \indexvariablenode}]
\bigl(\LLRsvector{\indexvariablenode}{\indexchecknode}\bigr).
$\\
To form the final decision, we compute for each \(\indexvariablenode \in \{1,\dots,n\}\)
\begin{equation}
  \LLR_{\indexvariablenode}^{(\zeta)}
  =
  \LLRcomponent{\indexvariablenode}^{(\zeta)}
  +
  \sum_{\substack{
      \indexchecknode \in \neighboringcn(\indexvariablenode) \\
      \langle \zeta,\;\stabilizermatrix{\indexchecknode,\indexvariablenode}\rangle = 1
  }}
  \messagechecknodes_{\arrowtoleft},
  \label{eq:message2}
\end{equation}
for all \(\zeta \in \mathbb{F}_{4} \setminus \{0\}\), and decide
$
\hat{\error}_{\indexvariablenode}
=
\arg\min_\zeta \LLR_{\indexvariablenode}^{(\zeta)}.
$
The iterations proceed until the syndrome is satisfied or until a maximum predefined number of iterations is reached.

\subsection{Quaternary neural belief-propagation}
\label{sec:neural-belief-propagation}

We use the \gls{nbp4} of \cite{miao2025quaternary}, which introduces the trainable weights on the edges of the Tanner graph, giving rise to the following message  update rules 
\begin{equation}
\messagechecknodes_{ \arrowtoleft}
=
(-1)^{\syndromecomponent_\indexchecknode}
\cdot
2  \tanh^{-1} \left(
\underset{\indexvariablenode' \in \neighboringvn(\indexchecknode) \setminus \{ \indexvariablenode\}}{\prod}
\tanh \frac{\weightvn[\indexvariablenode',\indexchecknode]^{(\indexiterations)} \cdot \beliefoperator[\indexvariablenode' \rightarrow \indexchecknode]}{2}
\right),
\label{eq: message cn weights}
\end{equation}
and
\begin{equation}
  \LLRsvectorcomponent{\indexvariablenode}{\indexchecknode}^{(\zeta)}
  \;=\;
  \weightvn[\indexvariablenode]^{(\indexiterations)} \LLRcomponent{\indexvariablenode}^{(\zeta)}
  \;+\;
  \sum_{\substack{
      \indexchecknode' \in \neighboringcn(\indexvariablenode)\setminus\{\indexchecknode\} \\
      \langle \zeta,\;\stabilizermatrix{\indexchecknode',\indexvariablenode}\rangle = 1
  }}
  \weightcn[\indexvariablenode,\indexchecknode']^{(\indexiterations)} \cdot \messagechecknodes_{\arrowtoleft'},
  \label{eq:message1weights}
\end{equation}
Here,  \( \indexiterations\) denotes the iteration index and
\( \weightvn[\indexvariablenode',\indexchecknode]^{(\indexiterations)} \),
\( \weightcn[\indexvariablenode,\indexchecknode']^{(\indexiterations)} \), and
\( \weightvn[\indexvariablenode]^{(\indexiterations)} \) are the real-valued trainable weights associated to the VN-to-CN messages, CN-to-VN messages, and channel log‑likelihood vectors, respectively.\\
For training, we adopt the  loss function of \cite{miao2025quaternary} that takes into account degeneracy:
\begin{equation}
    \mathcal{L}(\mathbf{\LLR};\error)
    =
    \sum_{\indexchecknode=1}^{2\numberqubits - \numberstabilizer}
    f\left(
      \sum_{\indexvariablenode=1}^{\numberqubits}
      P\bigl(\langle \error[\indexvariablenode] +\hat{\error}_{\indexvariablenode},
      \stabilizermatrix{\indexchecknode,\indexvariablenode}^{\perp}\rangle = 1
      \mid \boldsymbol{\syndromecomponent}\bigr)
    \right)
    \label{eq:loss}
\end{equation}
where
$
    P \left( \langle \hat{\error}_\indexvariablenode, \eta \rangle = 1 \mid \boldsymbol{\syndromecomponent} \right)
    =
    \left( 1 + \text{e}^{- \beliefoperator[\eta](\mathbf{\LLR}_\indexvariablenode)} \right)^{-1},
    \label{eq: hard decision step}
$
and $f(x)=|\sin(\pi x/2)|$ \cite{liu2019}. 
%, which approaches zero as \(x\) approaches any even integer (a ``soft modulo-2'' penalty) 
%Hence, the loss is accumulated over all rows $\boldsymbol{\stabilizer}_{\indexchecknode}^\perp$ of $\boldsymbol{\stabilizer}^{\perp}$; for each row, the inner sum in \eqref{eq:loss} aggregates the probabilities that the corresponding dual stabilizer is unsatisfied after applying the estimate \(\hat{\error}\).\\
The loss $\mathcal{L}$ is minimized when
$
    \langle\,\boldsymbol{\error[]}
  +\hat{\boldsymbol{\error[]}},
    \; \boldsymbol{\stabilizer}_\indexvariablenode^\perp
  \rangle = 0
  \label{eq:decision}
$
holds for every row $\indexvariablenode \in \{1, 2, \ldots, 2\numberqubits - \numberstabilizer\}$ of $\boldsymbol{\stabilizer}^{\perp}$. 
\subsection{Belief-matching}
The Belief-matching decoder couples binary \gls{bp} with \gls{mwpm} \cite{higgott2023improved,google2023suppressing} \footnote{Belief-matching, as originally introduced in \cite{higgott2023improved}, relies on binary BP. However, in this work we adopt quaternary \gls{bp}.}. In the first stage, \gls{bp} is run on the Tanner graph yielding posterior marginals for each error mechanism (variable node).  If these define a valid correction, it is accepted. Otherwise, the posteriors are used to \emph{reweight} the matching graph by decomposing each hyperedge into edges and redistributing probability mass, which is then converted into edge weights via a logarithmic map. 
Finally, weighted \gls{mwpm} is applied to obtain a syndrome-consistent correction. 
Unlike standard matching decoders, which use only prior error rates and ignore hyperedge correlations, this scheme injects posterior information into the matching step, allowing \gls{bp} to capture correlations such as those from $\boldsymbol{Y}$ errors and improving practical decoding performance.

\section{Proposed Decoders}
\label{sec:proposed decoder}

\subsection{Neural belief-matching}

We propose a novel two-stage decoder, \gls{nbm4}, that combines a neural belief‑propagation
pre‑processor (\gls{nbp4}) with minimum‑weight perfect matching (\gls{mwpm}).  
The method builds on the belief‑matching algorithm of \cite{higgott2023improved}
by replacing the binary BP step  with \gls{nbp4}
\footnote{Implementation taken from \cite{miao2025quaternary}
(https://github.com/kit-cel/Quantum-Neural-BP4-demo).}.

In particular, the decoder workflow is as follows. We first run \gls{nbp4}, which provides a log‑likelihood‑ratio vector $\mathbf{\Gamma}_{\indexvariablenode}$ for each variable node $v_{i}$. If \gls{nbp4} converges to a valid solution we terminate. 
Alternatively, i.e., if the output of \gls{nbp4} does not match the observed syndrome $\boldsymbol{s}$, the output of \gls{nbp4} is 
fed to the \gls{mwpm} algorithm to compute the final correction.
%used to reweight the edges in the matching graph 
%$\mathcal{G}$.

%
%This work introduces a hybrid decoding framework, \gls{nbm4}, which integrates \gls{nbp4} preprocessing\footnote{We use the \gls{nbp4} implementation from \cite{miao2025quaternary}. GitHub link: https://github.com/kit-cel/Quantum-Neural-BP4-demo} with \gls{mwpm}. Building upon the Belief-Matching algorithm \cite{higgott2023improved}, our approach extends the methodology to the quaternary domain by substituting the standard BP with \gls{nbp4}. The workflow in \gls{nbm4} proceeds as follows. Preprocessing with \gls{nbp4} involves executing neural decoders on the Tanner graph to estimate $\mathbf{\LLR}_{\indexvariablenode}$ for each variable  $v_\indexvariablenode$. If \gls{nbp4} fails to converge, due to unresolved syndromes or non-convergence, the computed   $\mathbf{\LLR}_{\indexvariablenode}$ values are used to reweight the edges in the matching graph  $\mathcal{G}$. 
%The reweighted graph $\mathcal{G}$ is then decoded using \gls{mwpm}. 
%This two-stage pipeline leverages the local inference power of \gls{nbp4} for most cases while retaining the global optimality guarantees of \gls{mwpm}. 
%The same workflow, considerations, and benefits apply to \gls{cbm4} with \gls{cnbp} replacing \gls{nbp4} in the preprocessing step. 
The key advantage of using \gls{nbp4} over traditional \gls{bp} lies in its ability to better handle degeneracy and loops in the Tanner graph, which are critical challenges in decoding topological codes. 
As a consequence, \gls{mwpm} needs to be called much less frequently, which yields a reduction of the average computational complexity. 

\subsection{Convolutional neural belief-propagation and weight-reuse strategy}

\label{subsec:ocnbp}

The \gls{cnbp} decoder introduced in this work is a \gls{cnn}-inspired architecture that leverages the toric code's periodic boundary conditions to implement weight-sharing, a fundamental feature of convolutional neural networks \cite{ketkar2021convolutional}. 
This approach is inspired by the weight-sharing framework introduced for toric codes in \cite{liu2019}, and by works on classical error correction \cite{wang2023ldpc,chen2021cyclically}.

\begin{figure}[t]
    \centering
    % riduco larghezza al 60% della colonna
    \includegraphics[width=1\columnwidth]{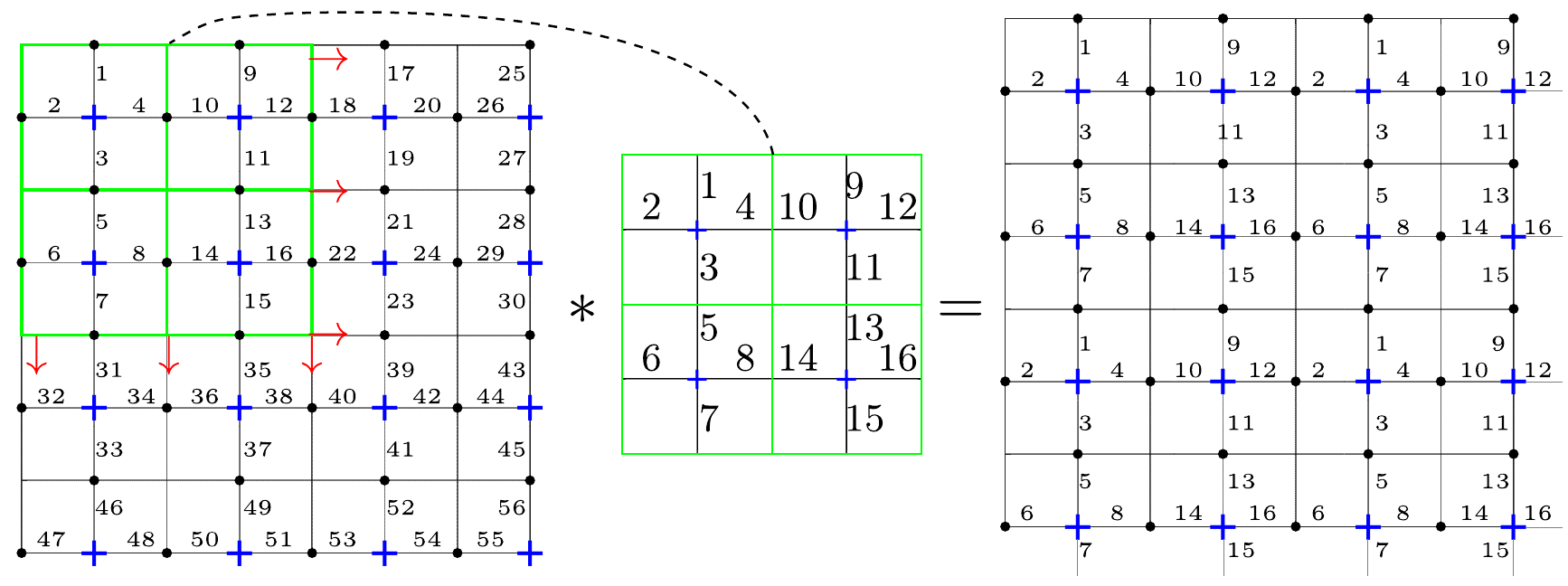}
    \caption{
Construction of \(\boldsymbol{\stabilizergroup} _{\text{type}}^{\text{\tiny CN}}\)  for $\boldsymbol{X}$‑stabilizers on a $d\!=\!4$ toric code. 
A $2\times2$ patch of $\boldsymbol{X}$‑stabilizers (and their edges) is tiled over the lattice, enforcing weight‑sharing and translation invariance; the $\boldsymbol{Z}$‑stabilizers are built analogously.
}
   % \caption{Graphical representation of \(\boldsymbol{\stabilizergroup} _{\text{type}}^{\text{\tiny CN}}\) construction for $\boldsymbol{X}$-Stabilizers in a toric code with $\codedistance=4$. A 2x2 patch containing a subset of $\boldsymbol{X}$-stabilizers and their corresponding edges is translated across the lattice. This process enforces weight-sharing by replicating the same patch structure at every position, ensuring translation invariance in the stabilizer configuration. The same considerations apply to  $\boldsymbol{Z}$-stabilizers.} 
 \label{fig:im2}
\vspace{-5mm}

\end{figure}

The defining feature of \gls{cnn} is the convolutional layer, with the convolutional operation defined as: $o_{\indexvariablenode,\indexchecknode} = h \left( \sum_{q=1}^D (\mathbf{K}_q * \mathbf{X}_q)(\indexvariablenode,\indexchecknode) + b \right)$
where \(\mathbf{K}\) is the kernel, \(\mathbf{X}\) is the input tensor, \(o_{\indexvariablenode,\indexchecknode}\) is the output of the layer (feature map) associated with the weights \(\weight_{\indexvariablenode,\indexchecknode}\), \(h\) is the activation function, \(b\) is the bias and $D$ the number of filters 
%\cite{lecun2002gradient}.\\
\cite{fukushima2007visual}.\\
To adapt this operation to our toric code, we define three kernel matrices: $\boldsymbol{\stabilizergroup}_{\text{type}}^{\text{\tiny VN}}$ and
        $\boldsymbol{\stabilizergroup}_{\text{type}}^{\text{\tiny CN}}$ for weights associated to  variable and check node messages (size $\numberstabilizer\times\numberqubits$) and 
 \(\boldsymbol{\stabilizergroup} _{\text{type}}^{\text{\tiny channel}}\)  for weights associated to  the channel log‑likelihood vectors (size $1\times\numberqubits$).
The kernel matrices are obtained by tiling a $2\times2$ stabilizer patch across the toric lattice (Fig.~\ref{fig:im2}).  The patch contains a fixed set of stabilizers, and all instances of the same stabilizer type share the same weight index.
 %\(\boldsymbol{\stabilizergroup} _{\text{type}}^{\text{\tiny LLR}}\)  is derived from categorizing the qubits in the lattice into 8 types based on their positions  within a $4 \times 4$ block (formed by tiling $2 \times 2 $ patches). This assignment ensures that qubits in equivalent geometrically positions across different patches share the same type.
\\%The LLR kernel groups the $\numberqubits$ qubits into eight geometric types within a $4$ super‑patch, guaranteeing that qubits in equivalent positions share a weight.\\

\begin{figure}[t]
    \centering
    % riduco larghezza al 60% della colonna
    \begin{overpic}[width=1\columnwidth]{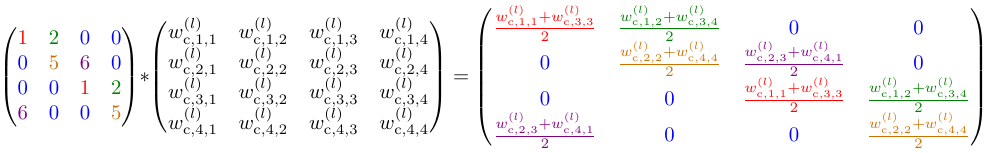}
        % coordinate (x,y) in percento rispetto all'immagine
        \put(27,16){\tiny$\boldsymbol{W}_{CN}$} 
   \put(5,16){\tiny $\boldsymbol{\stabilizergroup}_{\text{type}}^{\scaleto{\text{CN}}{2.5pt}}$}
    \put(72,16){\tiny$\boldsymbol{W}_{CN}^{*}$}
    \end{overpic}
    \caption{
Convolution in the convolutional NBP decoder: the check‑node weight tensor $\boldsymbol{W}_{CN}$ is filtered by 
\(\boldsymbol{\stabilizergroup} _{\text{type}}^{\text{\tiny CN}}\) to enforce weight-sharing, and average‑pooling over identical pivots produces the final tensor $\boldsymbol{W}_{CN}^{*}$.
}
 %   \caption{Example of the convolutional operation in our \gls{cnbp} decoder. 
%The check-node weight tensor $\boldsymbol{W}_{CN}$ is filtered through 
%\(\boldsymbol{\stabilizergroup} _{\text{type}}^{\text{\tiny CN}}\) , which indexes weights sharing the same value. 
%The final weight tensor $\boldsymbol{W}_{CN}^{*}$ is obtained via average pooling over weights corresponding to identical pivot values.}
 \label{fig:im3}
\end{figure}

Applying a kernel to the weight tensor of the code reproduces the CNN behaviour: the same filter is applied over the entire lattice (stride $2$, padding realized by the torus’s periodic boundaries).  The resulting activation map $o$ is passed to the next layer, and the bias $b$ is set to zero.  Average pooling over entries that share the same kernel index yields the weight‑shared tensors $\boldsymbol{W}_{\text{CN}}^{*}$, $\boldsymbol{W}_{\text{VN}}^{*}$, and $\boldsymbol{W}_{\text{channel}}^{*}$ (Fig.~\ref{fig:im3}).  The final fully‑connected layer is the NBP network described in Sec.~\ref{sec:neural-belief-propagation}.

\begin{figure}[t!]
    \centering
   %  \vspace{-1cm} 
    \begin{overpic}[width=0.8\columnwidth]{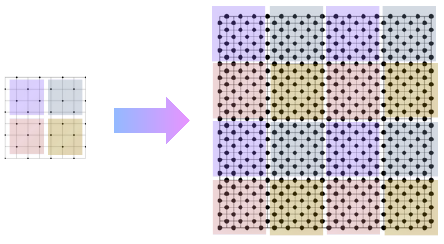}
        % Esempio di testo sovrapposto
         \put(15,14){\textcolor{quadBR}{$\boldsymbol{W}_{\codedistance\times\codedistance}^{4^{*}}$} }
         \put(0,41){\textcolor{quadTL}{$\boldsymbol{W}_{\codedistance\times\codedistance}^{1^{*}}$} }
         \put(0,14){\textcolor{quadBL}{$\boldsymbol{W}_{\codedistance\times\codedistance}^{3^{*}}$} }
         
         \put(15,41){\textcolor{quadTR}{$\boldsymbol{W}_{\codedistance\times\codedistance}^{2^{*}}$} }
   \put(55,-4){$(4<\codedistance) \times (4<\codedistance$)}
   
      \put(6,6){$4 \times 4$}
    
    \end{overpic}
    \vspace{0.5cm}
    \caption{
Weight‑reuse across code distances: the tensors $\boldsymbol{W}_{\codedistance\times\codedistance}^{n^{*}}$ for check nodes, variable nodes, and LLRs trained on a $d\!=\!4$ toric code are tiled to construct decoders for larger $d$, eliminating the need for costly retraining.
}
  %  \caption{Explanatory schema of weight-reuse in different code sizes. 
%The weight tensors $\boldsymbol{W}_{\codedistance\times\codedistance}^{n^{*}}$ for CNs, VNs, and LLRs, trained on a  $\codedistance=4$ toric code, can be reused for  codes with larger  $\codedistance$ via tiling the same patch structure. 
%This eliminates the need for retraining larger systems, which would otherwise require prohibitive computational resources.}
    \label{fig:im4}
\end{figure}
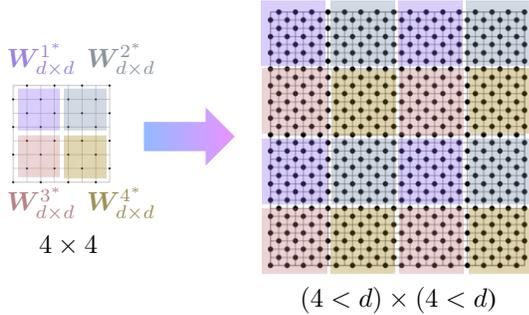

Weight‑sharing drastically reduces the number of trainable parameters, leading to faster training, lower memory usage, and enabling deeper networks %fukushima2007visual,
\cite{fukushima2007visual,lecun1989backpropagation,lecun2002gradient}. Because the decoder is built from a $2\times2$
patch, the weights learned for a code of distance $\codedistance$ can be tiled to decode any larger (or smaller) toric lattice with even dimension $L$ without retraining (Fig.~\ref{fig:im4}).
%, providing a modular and scalable solution for quantum error correction as the code size grows.

%\FloatBarrier

%\input{./tex_files/decoder/proposed_decoder.tex}
%\label{sec:proposed decoder}
%\subsection{Quaternary Convolutional Neural Belief Propagation }

%\subsection{Quaternary Neural  Belief Matching %and \\ Quaternary 
%Neural  Belief Find}

%\subsection{Training in Appendix}
%\input{./tex_files/results/training.tex}

\section{Numerical Results}

In this section we compare the performance of the proposed two-stage decoding algorithms to existing algorithms. We consider toric codes of distances $4, 6, 8$ and 10, and physical qubit error rates ranging from $0.015$ to $0.15$. 
The results are obtained through Monte Carlo simulations.

The figures in this section show 0.975 confidence intervals\footnote{{the confidence intervals in several figures are too small to be visible at the plot scale, rendering them indistinguishable from the curves. The raw data and detailed statistical analyses are available upon request.}} for each data point, computed using a negative binomial estimator \cite{mazzeo2011monte}.

%The figures in this section show confidence intervals for each data point assuming a negative binomial estimator and confidence intervals\footnote{{the confidence intervals in several figures are too small to be visible at the plot scale, rendering them indistinguishable from the curves. The raw data and detailed statistical analyses are available upon request.}
%in most of the figures, the confidence intervals are hard to real due to overlapping curves.
%} at a confidence level  of $0.975$ \cite{mazzeo2011monte}.

\subsection{NBP-Matching Results}
\label{subsec:nbm4 results}
Here, we evaluate the performance of belief-matching, \gls{nbm4}, and RNBP-matching (a variant  of \gls{onbp}) against the \gls{onbp} decoder proposed in \cite{miao2025quaternary} and the \gls{mwpm} decoder. The \gls{nbp4}`s training methodology  employed in this work is   the one proposed in \cite{miao2025quaternary}.
%For weighted \gls{uf}, we similarly compare Belief-Find, \gls{nbuf4}, and NBP$^{+}$-Find. 
The \gls{mwpm} results are generated using the Pymatching library \cite{higgott2022pymatching}.
%while the weighted \gls{uf} implementation leverages the LDPC library \cite{Roffe_LDPC_Python_tools_2022}.

\begin{figure}[t]
   \centering
    \includegraphics[width=0.9\columnwidth]{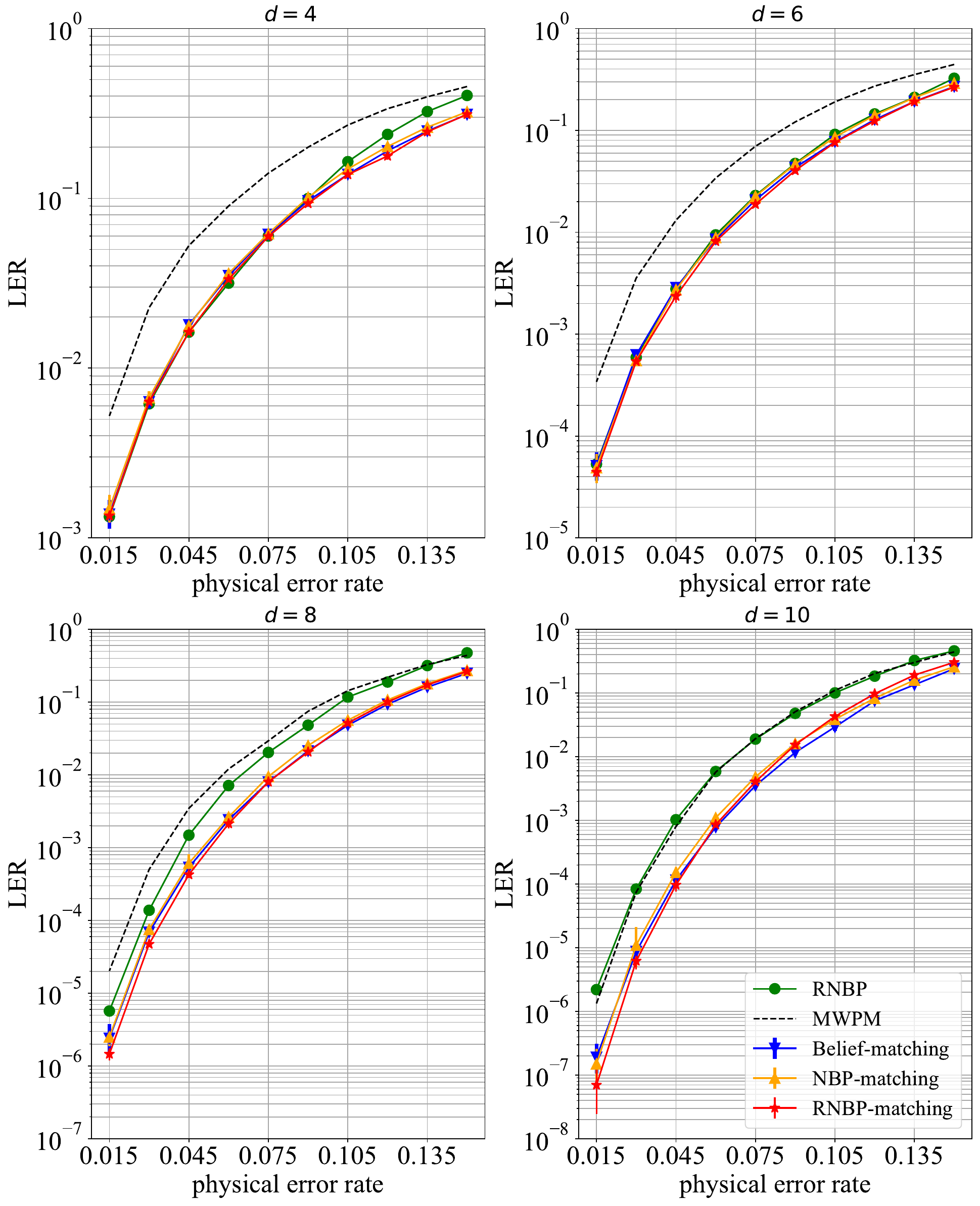}
    
    \caption{
\gls{ler} versus physical error rate for toric codes ($d\!=\!4,6,8,10$) using belief‑matching, \gls{nbm4}, and RNBP‑matching decoders. \gls{onbp} results are taken from \cite{miao2025quaternary}.
}
%    \caption{Comparison of the LER as a function of the physical error rate for toric codes with dimensions $\codedistance \in \{4,6,8,10\}$ utilizing the Belief-Matching, \gls{nbm4} and NBP$^{+}$-Matching decoders. The \gls{onbp} results are taken from \cite{miao2025quaternary}.}
    \label{fig:plot1}
\end{figure}

Fig. \ref{fig:plot1} shows the \gls{ler} as a function of the physical qubit error rate for the aforementioned decoders.
We can observe that all three two-step decoders (belief-matching, \gls{nbm4} and RNBP-matching) perform similarly. In particular, they outperform standalone \gls{mwpm} by  approximately one order of magnitude.  For code distances $d = 4,6$ \gls{onbp} achieves a similar performance to that of the two-stage decoders. 
However, for larger distances $d = 8$ and $d = 10$, the two-step decoders outperform \gls{onbp}.
%: at  $d = 10$, the improvement reaches one order of magnitude for both \gls{nbm4} and Belief-Matching, while NBP$^{+}$-Matching exceeds \gls{onbp} by more than one order of magnitude. 
%This trend demonstrates that the performance of our two-step decoders improves with increasing code size. Moreover, for $d =  10$, the NBP$^{+}$-Matching decoder performs slightly better then both \gls{nbm4} and Belief-Matching. \\
%Figure~\ref{fig:plot20} presents results for two-step decoders using weighted \gls{uf} as postprocessing. In this case, the two-step decoders again outperform \gls{uf} alone by one order of magnitude for $d = 4, 6, 8$ , and by two orders of magnitude at $d = 10$. %This indicates that the benefit of the two-step approach grows with code size.
%For $d = 4$ and $d = 6$, the \gls{nbuf4} and NBP$^{+}$-Find decoders perform comparably to \gls{onbp}. At $d = 8$,  \gls{nbuf4} and NBP$^{+}$-Find surpass Belief-Find and \gls{onbp}, with NBP$^{+}$-Find achieving better performance then \gls{nbuf4}. This underscores the effectiveness of training combined with an overcomplete check matrix. The impact of training becomes even more pronounced at $d = 10$: \gls{nbuf4} outperforms \gls{onbp} by one order of magnitude and  exceeds Belief-Find in a more marked way compared to the $d = 8$ case. Furthermore, the combination of training and an overcomplete check matrix enables NBP$^{+}$-Find to outperform \gls{onbp} by more than one order of magnitude and Belief-Find by one order of magnitude.\\

\begin{figure}[h!]
    \centering
    \includegraphics[width=0.9\columnwidth]{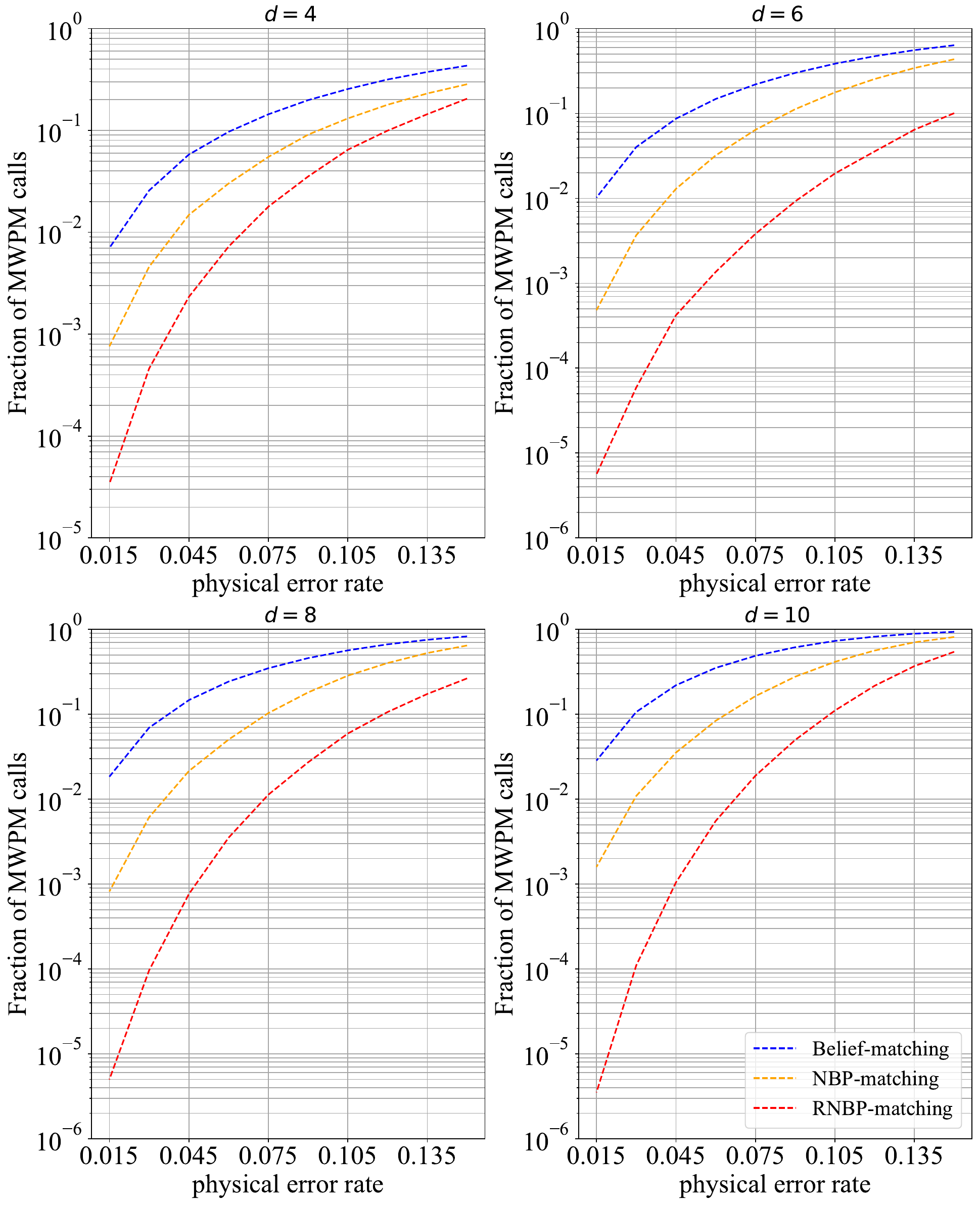}
   \caption{Comparison of the frequency of \gls{mwpm} algorithm calls as a postprocessing step, as a function of the physical error rate, with dimensions $\codedistance \in \{4,6,8,10\}$ utilizing the belief-matching, \gls{nbm4} and RNBP-matching decoders.}
    \label{fig:plot2}
\end{figure}

Fig. \ref{fig:plot2} shows the frequency of \gls{mwpm} algorithm invocations as a postprocessing step for the different two-stage decoders as a function of the physical qubit error rate. 
%The efficiency gains of training are evident in Fig. \ref{fig:plot2}, which shows the frequency of \gls{mwpm} algorithm invocations as a postprocessing step for the same code distances and decoders. 
The results show that the standard neural BP decoder (without using overcomplete matrices), \gls{nbm4}, reduces considerably the number of calls to \gls{mwpm}. 
The number of calls is further reduced when using the RNBP-matching decoder.
Thus, the use of overcomplete check matrices in the two-step decoder yields a significant reduction of the average computational cost of the decoder.

%%%%%%%%%%%%%%%%%%%%%%%%%%%%%%%%%%%%%%%%%%%%%%%%%%%%%%%%%%%%%%%%%%%%%%%%%%%%%%
%%%%%%%%%%%%%%%%%%%%%%%%%%%%%%%%%%%%%%%%%%%%%%%%%%%%%%%%%%%%%%%%%%%%%%%%%%%%%%
\subsection{Convolutional NBP and  $\textit{weight-reuse}$ strategy results}
\label{subsec:ocnbp results}

\begin{figure}[h!]
    \centering
    \includegraphics[width=0.9\columnwidth]{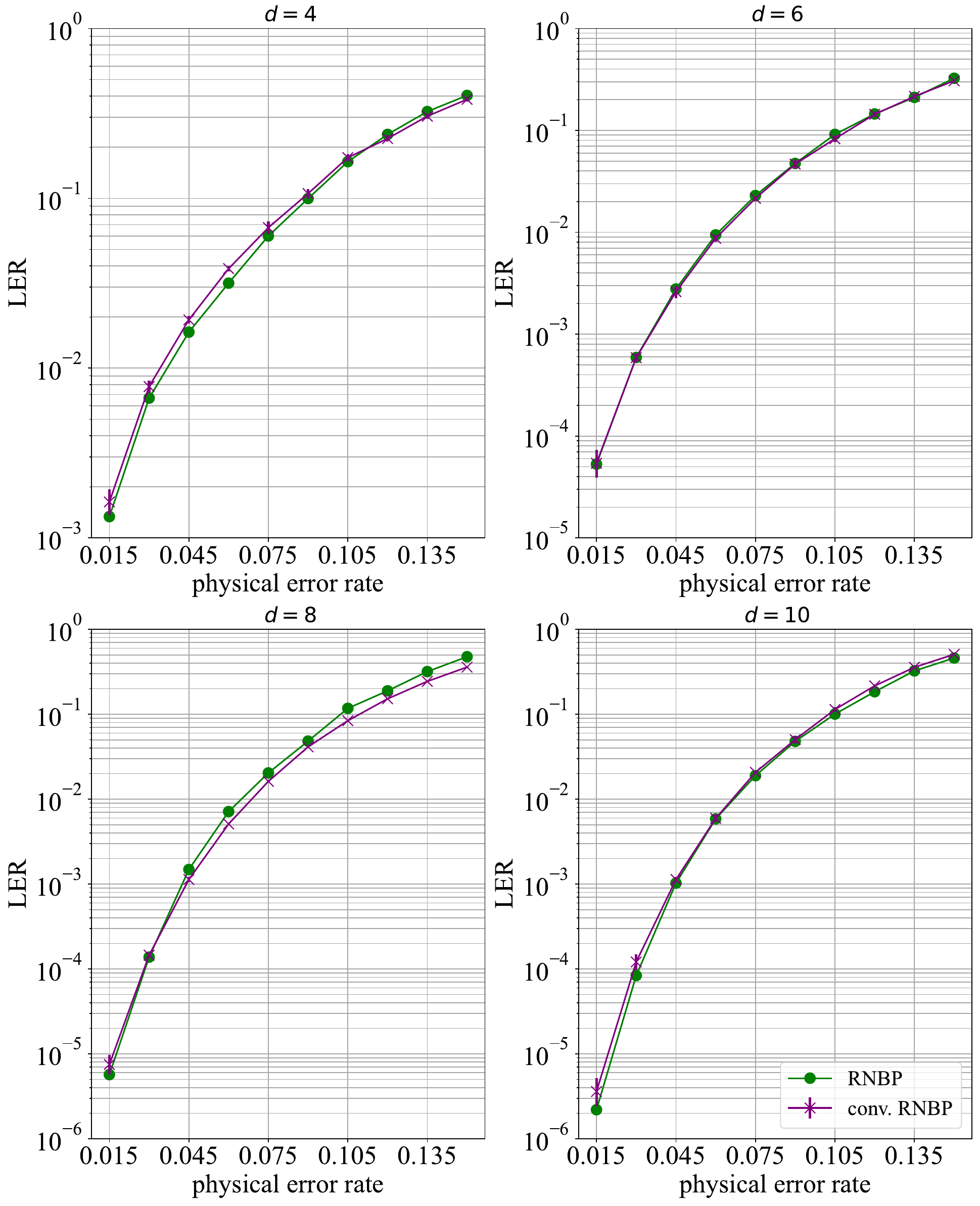}
    \caption{\gls{ler} versus physical error rate for toric codes ($d\!=\!4,6,8,10$) using the convolutional RNBP decoder.}
    \label{fig:plot3}
\end{figure}

\begin{figure}[ht]
    \centering
    \begin{subfigure}{0.7\linewidth}
      \includegraphics[width=\linewidth]{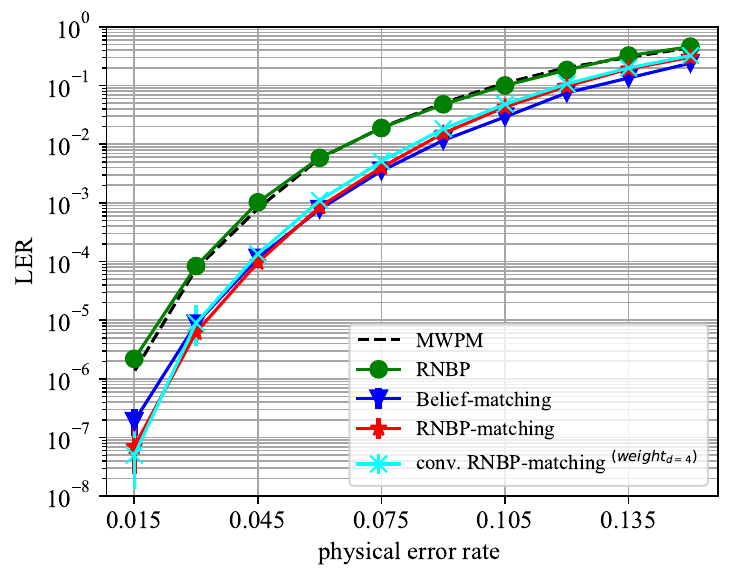}
        \caption{}
        \label{fig:subplot1}
    \end{subfigure}

    \begin{subfigure}{0.7\linewidth}
        \includegraphics[width=\linewidth]{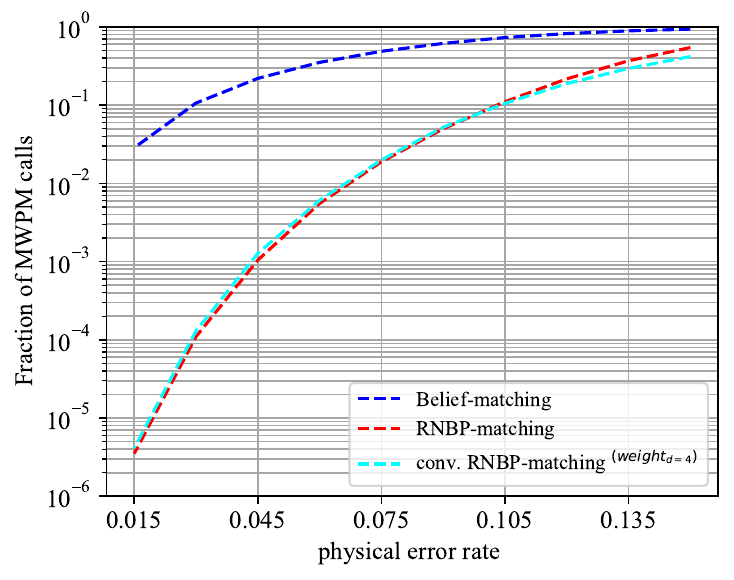}
        \caption{}
        \label{fig:subplot2}
    \end{subfigure}

    \caption{(a) \gls{ler} comparison for toric codes using convolutional RNBP and convolutional RNBP-matching decoders, with weights trained on 
$\codedistance=4$ applied to 
$\codedistance=10$, and \gls{onbp} results from \cite{miao2025quaternary}.\\
(b) Frequency of \gls{mwpm} postprocessing invocations for convolutional RNBP-matching decoders under the same conditions.}
    \label{fig:plots}
    \vspace{-8mm}

\end{figure}

We first evaluate the performance of the proposed \gls{ocnbp}. That is, we consider for the moment only the first-stage of the proposed decoders, without matching.
Fig. \ref{fig:plot3} shows the \gls{ler} as a function of the physical qubit error rate for   \gls{ocnbp} and \gls{onbp}, which is added for reference. 
The results demonstrate that \gls{cnn} decoder performs similarly to \gls{nn}-based one, despite having much fewer trainable weights.
We now evaluate the performance of the proposed weight-reuse strategy.
Fig. \ref{fig:plots}(a) shows the \gls{ler} as a function of the physical qubit error rate for different decoders for a toric code of distance $\codedistance=10$, including a \gls{ocnbp} decoder that reuses weights trained on a $\codedistance=4$ toric code.
As it can be observed, reusing the weights of $\codedistance=4$ yields essentially the same performance compared to using weights trained for $\codedistance=10$.

Fig. \ref{fig:plots}(b) shows the frequency with which \gls{mwpm} is called for 
belief-matching, RNBP-matching, and convolutional RNBP-matching reusing the weights obtained from  a $\codedistance=4$ toric code. As it can be observed,  reusing the weights obtained for  $\codedistance=4$ for a code with $\codedistance=10$, does not lead to more calls to \gls{mwpm}.

%Quaternary Convolutional Belief-Matching 
%decoder
%using an overcomplete check matrix
%(Convolutional NBP$^{+}$-Matching),
%which incorporates \gls{ocnbp} preprocessing trained on $\codedistance=4$. Additionally, we assess the performance of \gls{ocnbp} for $\codedistance=10$ using $\codedistance=4$-trained weights, benchmarking it against standard \gls{onbp} and \gls{mwpm} decoder. The \gls{ocnbp} decoder with $d=4$ weights achieves comparable performance to \gls{onbp} and \gls{mwpm} without significant degradation. Notably, Convolutional NBP$^{+}$-Matching outperforms both \gls{mwpm} and \gls{onbp} by an order of magnitude, matching the performance of NBP$^{+}$-Matching. 
%This approach avoids computationally prohibitive training for large codes by transferring weights from smaller ($\codedistance=4$) systems without accuracy loss. This efficiency gain is further demonstrated by the reduced frequency of \gls{mwpm} calls, as shown in Fig.~\ref{fig:plots}(b).

%\vspace{-5mm}

Thus, our results indicate that the  weight reuse strategy introduced in Section~\ref{subsec:ocnbp} does not result in loss in performance. This is quite remarkable, since it solves the scalability problem of training neural networks of increasing size.

\label{sec:numerical results}

\section{Conclusions}
\label{sec:conclusions}

This paper has presented two significant advancements in the decoding of toric codes for quantum error correction. First, we introduced the neural belief-matching decoder, which replaces the traditional belief-propagation stage with a neural belief-propagation (NBP) decoder. This modification yields a dramatic reduction in the average decoding complexity, decreasing the number of calls to the minimum-weight perfect matching (MWPM) decoder by up to four orders of magnitude for larger code distances (e.g., $10^{4}\times$ fewer calls for $d=8,10$). Despite this substantial reduction in complexity, the neural belief-matching decoder maintains the same logical error rate performance compared to standard belief-matching.
Second, we proposed a convolutional neural belief-propagation (convolutional NBP) decoder with a weight-reuse strategy that exploits the translational symmetry of the toric code. This approach enables training on a small code instance (e.g., 
$d=4$) and direct application to much larger codes (e.g., 
$d=10$) without any performance degradation. The weight-reuse strategy effectively addresses the scalability challenge of training neural decoders for larger quantum error correction codes, reducing the training burden while maintaining decoding quality.
These contributions are particularly significant in the context of fault-tolerant quantum computing, where real-time decoding is essential. The exponential slowdowns highlighted by Terhal's backlog argument \cite{terhal2015quantum} become increasingly problematic as quantum systems scale, making the reduction in reliance on computationally intensive \gls{mwpm} decoders critical for practical implementation. By minimizing the need for \gls{mwpm}, our approach ensures that decoding can keep pace with syndrome generation, enabling scalable quantum error correction systems.
The proposed techniques are not limited to toric codes and can be extended to other topological codes such as the surface code. The weight-reuse strategy shows particular promise for circuit-level noise models, where the Tanner graph complexity and training challenges are most pronounced. Future work will focus on extending these methods to more complex quantum error correction codes and exploring their application in practical quantum computing architectures.

\section{Acknowledgment}
We thank A. Fengler and S. Miao for useful discussions. 
%The work was made possible with the support of a scholarship from the German Academic Exchange Service (DAAD).

\bibliographystyle{IEEEtran}
\bibliography{IEEEabrv, ML_dec.bib}

\end{document}